\documentclass[prb,preprint]{revtex4-1} 

\usepackage{amsmath}  
\usepackage{amsfonts} 
\usepackage[dvipdfmx]{graphicx} 
\usepackage{comment}
\usepackage{bm}

\usepackage{lineno}

\newcommand{\bB}{\mathbf{B}}
\newcommand{\bE}{\mathbf{E}}

\newcommand{\bEC}{\bE_{\mathrm C}}
\newcommand{\bEI}{\bE_{\mathrm I}}

\newcommand{\be}{\mathbf{e}}

\newcommand{\beq}{\begin{equation}}

\newcommand{\bJ}{\mathbf{J}}

\newcommand{\br}{\mathbf{r}}
\newcommand{\bS}{\mathbf{S}}

\newcommand{\eeq}{\end{equation}}

\newcommand{\permitt}{\varepsilon_0}
\newcommand{\permea}{\mu_0}

\begin{document}

\title{Frequency Dependence of the Displacement Current Density}

\author{Toshio Hyodo}
\email{hyodot@post.kek.jp}
\affiliation{Slow Positron Facility, Institute of Materials Structure Science,
High Energy Accelerator Research Organization (KEK),
1-1 Oho, Tsukuba, Ibaraki, Japan 305-0801}

\begin{abstract}
In Maxwell's equations, the electric field can be expressed as the sum of the Coulombic field associated with the electric charge and the induced field associated with the time variation of the magnetic field from Faraday's law. The same holds true for the displacement current densities, which are the time-derivatives of the respective electric fields. In the case of an AC current of constant amplitude, the amplitude of the displacement current density associated with the Coulombic field is independent of the frequency. In contrast, the displacement current density associated with the induced field is very small at low frequencies, increasing initially as the square of the frequency, and ultimately becomes of the same order of magnitude as that of the Coulombic field.
\end{abstract}

\maketitle 

\section{Introduction} 

When an alternating current flows in a circuit containing a capacitor, an alternating magnetic field is present between the capacitor plates, as can be confirmed using a toroidal coil or a superconducting quantum interference device (SQUID). \cite{Thompson89, Rizzotto99, BartlettCorle85}
This geometry is often used to show that Amp\`{e}re's law must be supplemented to include a displacement current, becoming the Amp\`{e}re-Maxwell law. Then the introduced displacement current density is used to quantitatively explain the magnetic field existing between the capacitor plates. Usually the discussion of the capacitor's electric and magnetic fields ends here. However, the magnetic field is also alternating in time, and thus induces an electric field. This induced electric field also contributes to a displacement current, but, as will be shown, with an amplitude that increases with frequency.

Section II prepares readers to understand the frequency dependence by reminding them that the electric field $\bm{E}$ in Maxwell's equations consists of both Coulombic and induced electric fields. In Section III, the frequency dependence of the displacement current density is examined. Section IV presents a discussion and summary.

\section{Two kinds of electric fields} 
This section provides a brief review of concepts demonstrated in Ref. \onlinecite{Hyodo22}. The electric field $\bE(\br, t)$ in Maxwell's equations can be written as the sum of two quite different fields, the Coulombic field $\bEC(\br, t)$ and Faraday's induced field $\bEI(\br, t)$ as
\beq
\bE(\br, t) = \bEC(\br, t) + \bEI(\br, t).    \label{eq:E}
\eeq
Hereafter, the space and time coordinates $(\br,t)$ for the vector and scalar variables will be omitted unless they are necessary for clarity. The Coulombic electric field $\bEC$ satisfies Gauss's law 
\beq
\nabla \cdot \bEC = \frac{\rho}{\varepsilon_0},  \label{eq:Gausslaw1}
\eeq
where $\rho$ is the charge density and $\varepsilon_0$ is the vacuum electric permittivity. It is irrotational,
\beq
\nabla \times \bEC = 0.    \label{eq:irrot}
\eeq
The induced electric field $\bEI$ satisfies Faraday's law 
\beq
\nabla \times \bEI = - \frac{\partial \bB}{\partial t},      \label{eq:Faradlaw1}
\eeq
where $\bB$ is the magnetic field, and is divergence free,
\beq
\nabla \cdot \bEI = 0.      \label{eq:divfree}
\eeq
With Eqs. (\ref{eq:irrot}) and (\ref{eq:divfree}), $\bEC$ and $\bEI$ represent the unique Helmholtz decomposition of the electric field $\bE$.

A consequence of Eq. (\ref{eq:E}) is that the displacement current density in Maxwell's equations is the implicit sum of the corresponding displacement current densities as
\beq
\varepsilon_0\frac{\partial \bE}{\partial t} = \varepsilon_0\frac{\partial \bEC}{\partial t} + \varepsilon_0\frac{\partial \bEI}{\partial t}.  \label{eq:dcdsum}
\eeq
It follows that the law of charge conservation is, because of Eq. (\ref{eq:divfree}),
\beq
\varepsilon_0\nabla \cdot \frac{\partial \bE}{\partial t} = \varepsilon_0\nabla \cdot \frac{\partial \bEC}{\partial t} = \frac{\partial \rho}{\partial t} =- \nabla \cdot \bJ,
\label{eq:chargecons1}
\eeq
and that the Amp\`{e}re-Maxwell law is
\beq
\nabla \times \bB =  \mu_0 \left(\bJ +  \varepsilon_0\frac{\partial \bE}{\partial t} \right) = \mu_0 \left(\bJ +  \varepsilon_0 \frac{\partial \bEC}{\partial t} +  \varepsilon_0 \frac{\partial \bEI}{\partial t} \right).
 \label{eq:AmpMaxlaw1} 
\eeq

\section{Frequency-dependence of the Displacement Current Density}
Now let's examine the time dependence of the electric field and the displacement current density in a circuit containing a capacitor. When an AC current $I(t) = I_0 \sin \omega t$ ﬂows in the circuit, all the ﬁelds vary with $\omega$. For simplicity we assume that the capacitor plates are circular with a diameter of $L$ (a radius of $r = L/2$), and the distance between the plates is small enough to allow fringing fields to be ignored.

In the quasistatic regime, the typical size $L$ of the system is much smaller than the distance that light travels during a typical period of time variation,
\beq
 L \ll \frac{c}{\omega},     \label{eq:qstat}
\eeq
Here $c$ is the speed of light and $\omega$ is the angular frequency of the time variation.  In this case the fields at any moment are analyzed by the static electromagnetic theory.

We denote the amplitudes of the fields $\bEC$, $\bEI$, $\permitt \partial \bEC /\partial t$, and $\permitt \partial \bEI /\partial t$ as $E_\mathrm{C0}$, $E_\mathrm{I0}$, $J_\mathrm{C0}$, and $J_\mathrm{I0}$, respectively.

\begin{figure}[h!]
\centering
\includegraphics[width=8cm]{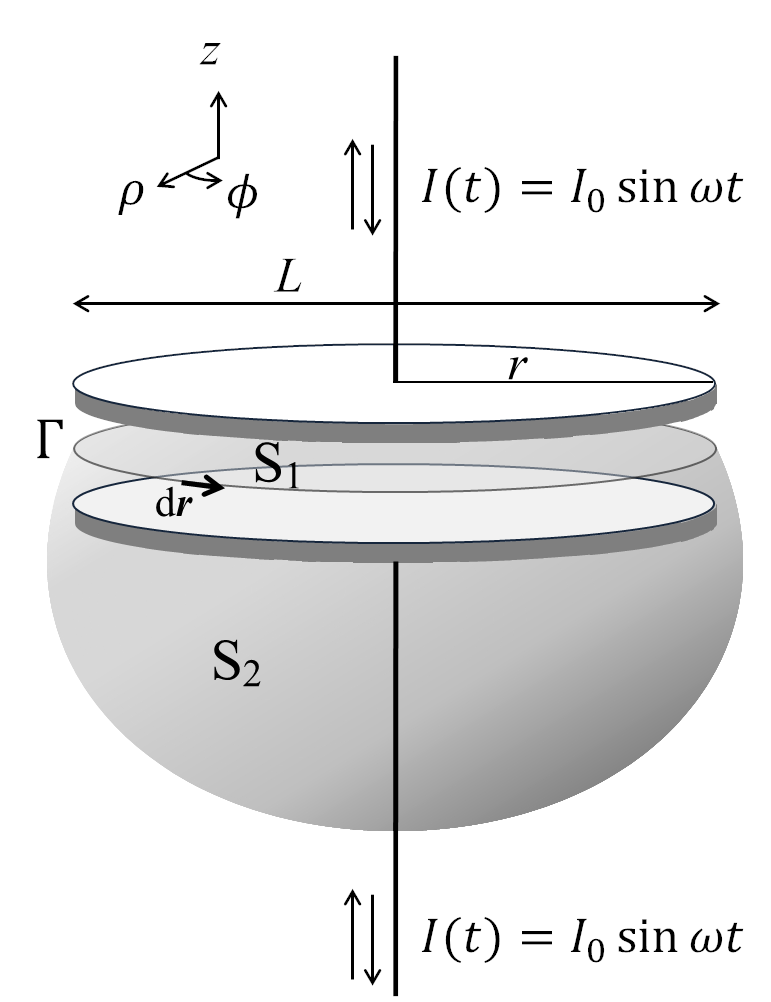}
\caption{An AC current $I(t)= I_0 \sin \omega t$ flowing through a capacitor of plates of diameter $L$ (radius $r$). A closed surface $\mathrm{S}_1 + \mathrm{S}_2$ surrounds a capacitor plate. The surface $\mathrm{S}_1$ is a disk bordered by a circle $\Gamma$ of diameter $L$ (radius $r$) lying in the middle of the capacitor plates, and $\mathrm{S}_2$ is another surface bordered by $\Gamma$ but around the back of the plate.  The line element vector $d\br$ that will be used in Faraday's law is indicated. The directions of the coordinates ($\rho, \phi, z$ ) in the cylindrical system are also shown.}
\label{fig:Fig1}
\end{figure}

First, $\permitt \partial \bEC /\partial t$ associated with $\bEC$, which is generated by the charge density on the plates,  is estimated using the integral form of the charge conservation law (\ref{eq:chargecons1}) as
\beq
 \permitt \oint_\mathrm{S} \frac{\partial \bEC}{\partial t} \cdot d \bS = -\oint_\mathrm{S} \bJ \cdot d \bS,     \label{eq:ccEC}
\eeq
where S is a closed surface. In the surface integral, the surface element vector $d\bS$ points outwards everywhere on S. We apply this to the closed surface $\mathrm{S} = \mathrm{S}_1 + \mathrm{S}_2$ enclosing one of the capacitor plates as in Fig. \ref{fig:Fig1}, where $\mathrm{S}_1$  is a disk bordered by a circle $\Gamma$ of radius  $r$ in the middle of the space between the capacitor plates, and $\mathrm{S}_2$ is another surface bordered by $\Gamma$ but around the back of the plate.  Then the right-hand side of Eq. (\ref{eq:ccEC}) is
\beq
 -\oint_\mathrm{S} \bJ \cdot d \bS = I,
\eeq
where the current flowing upward in the leads is taken to be in the positive direction. The left-hand side is, in the approximation of neglecting the fringing effects,
\beq
  \permitt \oint_\mathrm{S} \frac{\partial \bEC}{\partial t} \cdot d \bS 
  \sim \permitt \frac{d}{dt} \int_{\mathrm{S}_1}  \bEC \cdot d \bS
  = \pi r^2 \permitt \frac{d\overline{E_\mathrm{C}^{\mathrm{S}_1}}}{dt},   
\eeq
where $\overline{E_\mathrm{C}^{\mathrm{S}_1}}$ denotes the spatial average of $E_\mathrm{C}$ over the surface $\mathrm{S}_1$. We use similar notations also for the average of other quantities. Then the displacement current density associated with the Coulombic electric field is estimated as
\beq
  \permitt \frac{d\overline{E_\mathrm{C}^{\mathrm{S}_1}}}{dt} \sim \frac{I}{\pi r^2},  
   \label{eq:Cdcd}
\eeq
which is in phase with the current, and its amplitude is estimated as
\beq
 \overline{J_{\mathrm C0}^{\mathrm{S}_1}} \sim \frac{I_0}{\pi r^2}.     \label{eq:JC0}
\eeq
The amplitude $J_{\mathrm C0}$  is independent of $\omega$. Since the charge density on the capacitor plates and hence the electric field $\bEC$ produced by it is proportional to the time integral of the current, the amplitude of the displacement current density, which is the differential with respect to time of this field, does not depend on $\omega$.

Next, in order to estimate $\bB$ in the area between the capacitor plates, where conduction current is absent, we apply the integral form of the Amp\`{e}re-Maxwell law (\ref{eq:AmpMaxlaw1})  to the closed circle $\mathrm{\Gamma}_\rho$ of radius $\rho$ on the surface $\mathrm{S}_1$ in Fig. \ref{fig:Fig1} and the disk $\mathrm{S}_\rho$ bordered by $\mathrm{\Gamma}_\rho$ as
\beq
 \oint_{\mathrm{\Gamma}_\rho} \bB \cdot d \br 
 = \permea \int_{\mathrm{S}_\rho} \permitt \frac{\partial \bE}{\partial t} \cdot d \bS     
 = \permea \permitt  \frac{d}{dt}\int_{\mathrm{S}_\rho} \bE \cdot d \bS.      \label{eq:AmpMaxlaw2}
\eeq
In the quasistatic approximation, $\bB$ can be calculated by assuming $\bE = \bEC$. Note that Eq. (\ref{eq:AmpMaxlaw2})  does not imply that the displacement current  generates the magnetic field; it simply reflects the relationship that must exist between the fields. It will be shown later in this section that neglecting the contribution of induced electric field $\bEI$ is a good approximation in the quasistatic regime. We then have

\beq
 2\pi \rho B(\rho) \sim \pi \rho^2 \permea \permitt \frac{d\overline{E_\mathrm{C}^{\mathrm{S}_\rho}}}{d t} \sim \pi \rho^2 \permea \permitt \frac{d\overline{E_\mathrm{C}^{\mathrm{S}_1}}}{d t}  \mspace{18mu} \mathrm{for} \mspace{12mu}  \rho \leq r \label{eq:Bestm}
\eeq
since $\bEC$ is almost uniform in this area. Therefore, the magnetic field in the area between the capacitor plates is estimated as a function of radius $\rho$ to be
\beq
  \bB(\rho) \sim \frac{\permea \permitt}{2} \frac{d\overline{E_\mathrm{C}^{\mathrm{S}_1}}}{d t}  \rho \be_\phi  \mspace{18mu}  \mathrm{for} \mspace{12mu}  \rho \leq r. \label{eq:Brho}
\eeq
While $\bEC$ is confined in the area between the plates except for the fringing field, $\bB$ extends to the outside as
\beq
  \bB(\rho) \sim \frac{\permea \permitt}{2} \frac{d\overline{E_\mathrm{C}^{\mathrm{S}_1}}}{d t} \frac{r^2}{\rho} \be_\phi \mspace{18mu} \mathrm{for} \mspace{12mu}  \rho>r.
\label{eq:BrhoOut}
\eeq
The argument up to this point is essentially the same as is found in many textbooks.

\begin{figure}[h!]
\centering
\includegraphics[width=8cm]{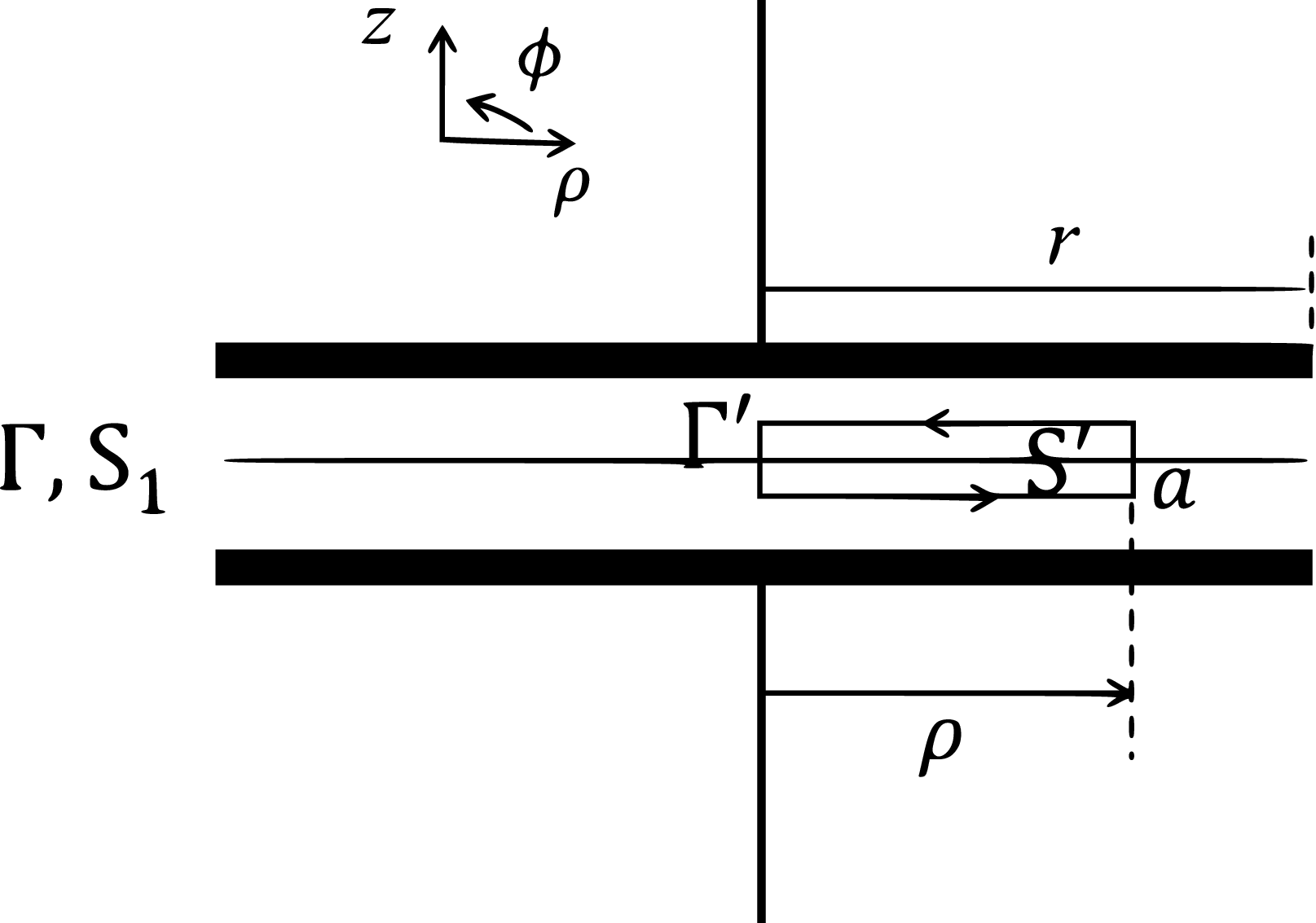}
\caption{Side view of the capacitor in Fig. 1. A vertical rectangular closed loop $\Gamma'$ crossing the surface $\mathrm{S}_1$ has horizontal sides of length $\rho$ and vertical sides of length $a$. The surface $\mathrm{S'}$ is bordered by $\Gamma'$.}
\label{fig:Fig2}
\end{figure}

Since $\bB$ varies with the angular frequency $\omega$, it generates an induced electric field $\bEI$. The field $\bEI$ is estimated using the integral form of Faraday's law applied to the surface $\mathrm{S'}$ bordered by the rectangular closed loop $\Gamma'$ crossing the surface $\mathrm{S}_1$  as shown in Fig. \ref{fig:Fig2},
\beq
 \oint_{\mathrm{\Gamma}'} \bEI \cdot d \br 
 = - \int_{\mathrm{S}'} \frac{\partial \bB}{\partial t} \cdot d \bS.     
 \label{eq:AmpMaxlaw3}
\eeq
Here the direction of the line element vector $d\br$ is set to be anticlockwise and the corresponding direction of $d\bS$ is pointing out of the page (from the back of the paper to the front). 
The loop $\Gamma'$ has horizontal sides of length $\rho$ and vertical sides of small length $a$ one of which is on the $z$ axis. Since $\bEI$ has only a $z$ component on $\mathrm{S}_1$ (see Appendix), the contributions of the line integral along the horizontal sides are zero. Thus the left-hand side of Eq. (\ref{eq:AmpMaxlaw3}) is
\beq
 \oint_{\mathrm{\Gamma}'} \bEI \cdot d \br \sim aE_\mathrm{I}(\rho),
\eeq
where we assumed  $E_\mathrm{I} (0)=0$ because $\bB$  at $\rho=0$ stays zero all the time. The right-hand side is, noting Eq. (\ref{eq:Brho}) and $\be_\phi \cdot d\bS \sim -1$ on $\mathrm{S}'$,
\beq
 - \int_{\mathrm{S}'} \frac{\partial \bB}{\partial t} \cdot d \bS
  \sim \frac{\permea \permitt}{2} \frac{d^2\overline{E_\mathrm{C}^{\mathrm{S}_1}}}{d t^2} a \int_0^\rho xdx = \frac{\permea \permitt}{4} \frac{d^2 \overline{E_\mathrm{C}^{\mathrm{S}_1}}}{d t^2} a\rho^2. 
\eeq
Thus
\beq
  E_I(\rho) \sim \frac{\permea \permitt}{4} \frac{d^2 \overline{E_\mathrm{C}^{\mathrm{S}_1}}}{d t^2} \rho^2.     \label{eq:EIrho1}
\eeq
It is possible to express this in terms of $B(\rho)$ by referring to Eq. (\ref{eq:Brho}) as
\beq
  E_I(\rho) \sim \frac{1}{2} \frac{dB(\rho)}{d t} \rho.    \label{eq:EIrho2}
\eeq

Now we are ready to compare the orders of magnitude of $\permitt \partial \bEC /\partial t $ and $\permitt \partial \bEI /\partial t $ by using the integral form of the Amp\`{e}re-Maxwell law, Eq. (\ref{eq:AmpMaxlaw1}), applied to $\Gamma$ and $\mathrm{S}_1$ in Fig. \ref{fig:Fig1}, 
\beq
 \oint_{\mathrm{\Gamma}} \bB \cdot d \br 
 = \permea \int_{\mathrm{S}_1} \left( \permitt \frac{\partial \bEC}{\partial t} + \permitt \frac{\partial \bEI}{\partial t} \right)\cdot d \bS.     
\label{eq:AmpMaxlaw4}
\eeq
Differentiating both sides by $t$ yields
\beq
 \oint_{\mathrm{\Gamma}} \frac{\partial \bB}{\partial t} \cdot d \br 
 = \permea \int_{\mathrm{S}_1} \left( \permitt \frac{\partial^2 \bEC}{\partial t^2} + \permitt \frac{\partial^2 \bEI}{\partial t^2} \right)\cdot d \bS. 
    \label{eq:diffAmpMaxlaw3}
\eeq
The left-hand side is, considering Eq. (\ref{eq:EIrho2}) for $\rho = r$, 
\beq
 \oint_{\mathrm{\Gamma}} \frac{\partial \bB}{\partial t} \cdot d \br 
  \sim 2 \pi r  \frac{dB(r)}{d t} = 4 \pi E_I(r).
\eeq
It is possible to express $ E_I(r)$ in terms of $\overline{E_\mathrm{I}^{\mathrm{S}_1}}$. Using Eq. (\ref{eq:EIrho1}) yields
\beq
 \overline{E_\mathrm{I}^{\mathrm{S}_1}} 
  = \frac{1}{\pi r^2} \int_{\mathrm{S}_1} \bEI \cdot d\bS 
  = \frac{1}{\pi r^2} \int_0^r  E_I(\rho) 2\pi \rho d\rho
  \sim \frac{\permea \permitt}{2r^2} \frac{d^2 \overline{E_\mathrm{C}^{\mathrm{S}_1}}}{d t^2}     \int_0^r \rho^3 d\rho
  = \frac{\permea \permitt}{8} \frac{d^2 \overline{E_\mathrm{C}^{\mathrm{S}_1}}}{d t^2} r^2.
\eeq
Then, comparing the right-hand side with Eq. (\ref{eq:EIrho1}) for $\rho = r$, we have
\beq
  E_I(r) \sim 2\overline{E_\mathrm{I}^{\mathrm{S}_1}}
\eeq
Therefore, the left-hand side of Eq. (\ref{eq:diffAmpMaxlaw3}) is
\beq
  \oint_{\mathrm{\Gamma}} \frac{\partial \bB}{\partial t} \cdot d \br  \sim 8 \pi \overline{E_\mathrm{I}^{\mathrm{S}_1}}       \label{eq:lhs}
\eeq
The right-hand side of Eq. (\ref{eq:diffAmpMaxlaw3}) is, by referring to Eq. (\ref{eq:Cdcd}) and noting $\permitt \permea = 1/ c $ and $d^2 \overline{E_\mathrm{I}^{\mathrm{S}_1}}/d t^2 = -\omega^2 \overline{E_\mathrm{I}^{\mathrm{S}_1}}$, 
\beq
 \permea \permitt \int_{\mathrm{S}_1} \left(  \frac{\partial^2 \bEC}{\partial t^2} + \permitt \frac{\partial^2 \bEI}{\partial t^2} \right)\cdot d \bS 
 =  \pi r^2 \permea \permitt \left(  \frac{d^2 \overline{E_\mathrm{C}^{\mathrm{S}_1}}}{d t^2} + \frac{d^2 \overline{E_\mathrm{I}^{\mathrm{S}_1}}}{dl t^2} \right)
  \sim \permea \frac{d I}{d t} - \pi \left( \frac{\omega r}{c} \right)^2 \overline{E_\mathrm{I}^{\mathrm{S}_1}}      \label{eq:rhs}
\eeq
From Eqs. (\ref{eq:lhs}) and (\ref{eq:rhs}), Eq. (\ref{eq:diffAmpMaxlaw3}) is reduced  to 
\beq
   8 \pi \overline{E_\mathrm{I}^{\mathrm{S}_1}}  \sim \permea \frac{d I}{d t} - \pi \left( \frac{\omega r}{c} \right)^2 \overline{E_\mathrm{I}^{\mathrm{S}_1}}   
\eeq
and thus 
\beq
   \overline{E_\mathrm{I}^{\mathrm{S}_1}}  \sim \frac{ \permea d I / d t}{\pi [8+ (\omega r/c)^2] }   
\eeq
Differentiating both sides by $t$ yields
\beq
   \permitt \frac{d\overline{E_\mathrm{I}^{\mathrm{S}_1}}}{dt} 
   \sim \frac{ \permitt \permea d^2 I / d t^2}{\pi [8+ (\omega r/c)^2]}   
   = - \frac{ \omega^2 I }{\pi c^2 [8+ (\omega r/c)^2] }.       \label{eq:dcdI}
\eeq
indicating that $\permitt \partial \bEI /\partial t$ and $I$ are in opposite phases. For the amplitude we have
\beq
 \overline{J_\mathrm{I0}^{\mathrm{S}_1}} \sim \frac{ \omega^2 I_0 }{\pi c^2 [8+ (\omega r/c)^2] } 
    \label{eq:JI0}
\eeq

 From Eqs. (\ref{eq:JC0}) and (\ref{eq:JI0}), the ratio of the amplitudes of the coexisting displacement current densities is
\beq
 \frac{\overline{J_\mathrm{I0}^{\mathrm{S}_1}}}{\overline{J_\mathrm{C0}^{\mathrm{S}_1}}} \sim \frac{(\omega r/c)^2 }{8+ (\omega r/c)^2 }.      \label{eq:ratioaf}
\eeq
At low frequencies over which Eq. (\ref{eq:qstat}) holds, the ratio simplifies to 
\beq
    \frac{\overline{J_\mathrm{I0}^{\mathrm{S}_1}}}{\overline{J_\mathrm{C0}^{\mathrm{S}_1}}} \sim \frac{1}{8} \left(\frac{\omega r}{c}\right)^2, 
\eeq
indicating that at low frequencies $\permitt \partial \bEC /\partial t$ is far more dominant than $\permitt \partial \bEI /\partial t$ with the latter increasing as $\omega^2$.

We may consider the frequency dependence of $\permitt \partial \bEI /\partial t$ in some more detail using Eqs. (\ref{eq:dcdI}) and (\ref{eq:JI0}). The validity of using them beyond the quasistatic approximation is discussed later in Section IV. Eqs. (\ref{eq:dcdI}) and (\ref{eq:JI0}) show that $\permitt \partial \bEI /\partial t$ continues to increase  as  $\omega^2$. The rate of increase slows down around $\omega = c/r$, where $\overline{J_\mathrm{I0}^{\mathrm{S}_1}} \sim \overline{J_\mathrm{C0}^{\mathrm{S}_1}}/9$.
When the frequency is so high that $r\omega/c \gg 1$, we have from Eq. (\ref{eq:JI0})
\beq
 \overline{J_\mathrm{I0}^{\mathrm{S}_1}} \sim \frac{I_0}{\pi r^2} = \overline{J_\mathrm{C0}^{\mathrm{S}_1}} ,       
\eeq
indicating that $\overline{J_\mathrm{I0}^{\mathrm{S}_1}}$ approaches $\overline{J_\mathrm{C0}^{\mathrm{S}_1}}$ and does not depend on the frequency any more. 

\begin{figure}[h!]
\centering
\includegraphics[clip=false, width=10cm]{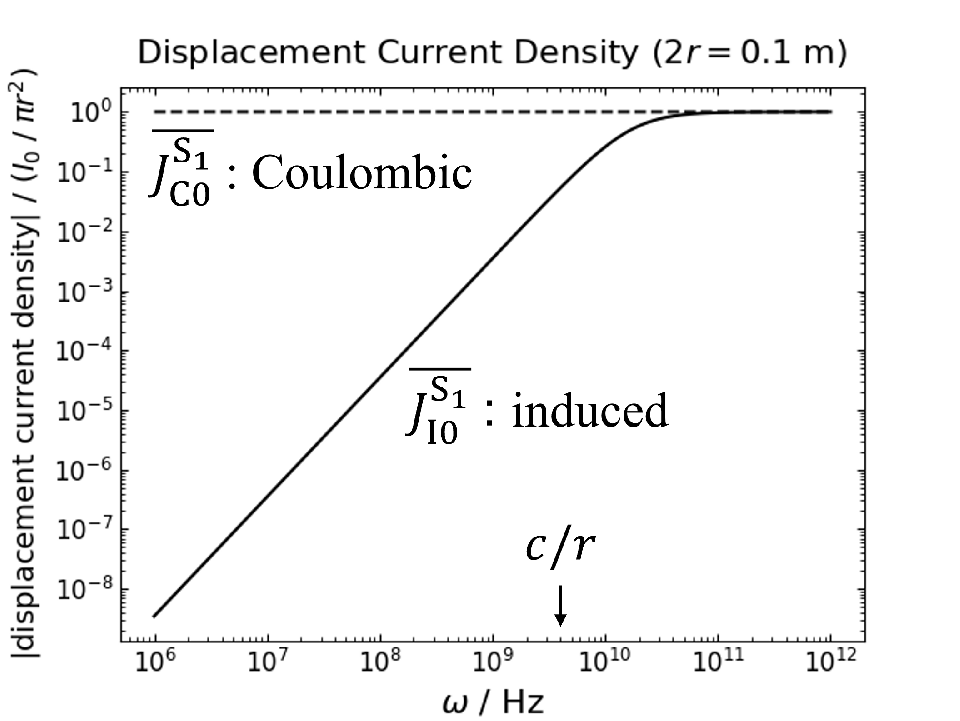}
\caption{Frequency dependence of the approximate averaged amplitudes $\overline{J_\mathrm{C0}^{\mathrm{S}_1}}$ and $\overline{J_\mathrm{I0}^{\mathrm{S}_1}}$ for a capacitor of diameter 10 cm.}
\label{fig:Fig3}
\end{figure}

Figure \ref{fig:Fig3} shows the frequency dependence of $\overline{J_\mathrm{C0}^{\mathrm{S}_1}}$ and $\overline{J_\mathrm{I0}^{\mathrm{S}_1}}$ in the case of $r=5$ cm. Here the ratio given by Eq.  (\ref{eq:ratioaf}) is less than $10^{-7}$ for a frequency $f = \omega /2 \pi <$ 1 MHz ($\omega < $6 MHz), which means that $\permitt \partial \bEC /\partial t$ is far more dominant. The amplitude $J_{\mathrm{I}0}$ increases in proportion to  $\omega^2$, then slows down as the frequency is approaching the limit of the quasistatic approximation, and then ultimately approaches $\overline{J_\mathrm{C0}^{\mathrm{S}_1}}$.

It is also interesting to note that the gate of a transistor in a modern central processing unit (CPU), typically with a size as small as $L\sim10$ nm and a frequency as high as $\sim 5$ GHz, operates in the quasistatic regime where $\permitt \partial \bEC /\partial t$ is dominant since $(\omega L/c)^2$  is of the order of $10^{-12}$.

\section{Discussion and Conclusion}
Introductory textbooks on electricity and magnetism usually present the displacement current density as existing between the plates of a capacitor, followed by a demonstration of how to calculate the magnetic field there by using the Amp\`{e}re-Maxwell law. Later, the wave equation for electromagnetic waves in vacuum is derived by coupling the Amp\`{e}re-Maxwell law and Faraday's law. While these presentations and mathematical treatments are not wrong, qualitative explanations that associate the displacement current density between the capacitor plates in the quasistatic regime with that involved in electromagnetic waves are found not to be correct. 

The treatment of time-dependent circuits by quasi-stationary approximation has two key characteristics. First, the transmission of information, which occurs at the speed of light, is regarded as virtually instantaneous. Second, the displacement current density of the induced electric field is ignored. While the former is always explicitly stated, the latter is only assumed implicitly. The present article has presented the rationale for the latter  by examining the frequency dependence of the displacement current densities.

The integral forms of the law of charge conservation and Maxwell's equations have been employed to estimate the amplitudes of the two different displacement current densities. Since the differential forms of these equations are locally correct at any space-time point $(\br, t)$ in an inertial frame of reference, they are always correct regardless of the speed of the time variation of the fields. Their integral forms are also correct even in a very rapidly changing time-dependent field regardless of the size of the integration domain. This is because they are not the integral solutions of the differential forms treated as partial differential equations, but are derived by simply applying vector analysis to the differential forms with the time variable fixed at the same value throughout the whole space of the observer's frame of reference. They indicate the correlations among the fields at the particular moment regardless of the retarded time of the source of each field at each point. Therefore the argument in Section III covers the whole frequency range.

Fig. 3 indicates that the composition of the displacement current density changes with the frequency of the current in the circuit. It shows that while the spatial average of the amplitude of $\permitt \partial \bEC / \partial t$ (dashed line) is independent of the frequency, that of $\permitt \partial \bEI / \partial t$ (solid curve) grows from an extremely small value to being comparable with the former. Although the components have opposite phases, this does not imply that they will eventually totally cancel each other. While $\permitt \partial \bEC / \partial t$ is almost uniformly confined in the area between the capacitor plates, $\permitt \partial \bEI / \partial t$ is not; $\permitt \partial \bEI / \partial t$ and  $J_{\mathrm{I}0}(\rho)$ must be zero at the center of the area between the capacitor plates and becomes large as the distance from the center, $\rho$, increases. Thus, $J_{\mathrm{I}0}(\rho)$ must be greater than  $J_{\mathrm{C}0}(\rho)$ in some outer range of $\rho$. Moreover, $\permitt \partial \bEI / \partial t$ extends to the outside since $\bB$ does so as indicated by Eq. (\ref{eq:BrhoOut}).

It is to be noted that only $\permitt \partial \bEI / \partial t$ is involved in the electromagnetic wave. This can be understood by noticing that $\permitt \partial \bEC / \partial t$ vanishes in the first step of the derivation of the wave equation, namely, taking the curl of both sides of the Amp\`{e}re-Maxwell's law.\cite{Hyodo22}

In summary, the present article has confirmed that the displacement current density is not a single entity; it is composed of two parts ---one associated with the Coulombic electric field and the other with the induced electric field--- with very different physical characteristics and effects. Then it has revealed that the operating frequency determines which is dominant or of equal magnitude. Clarifying this distinction will help prevent misconceptions in both educational and theoretical interpretation.

\section*{Appendix: Direction of $\bEI$ in the middle between the capacitor plates}
\renewcommand{\theequation}{A\arabic{equation}}
\setcounter{equation}{0}
\renewcommand{\thefigure}{A\arabic{figure}}
\setcounter{figure}{0}

\begin{figure}[h!]
\centering
\includegraphics[clip=false, width=8cm]{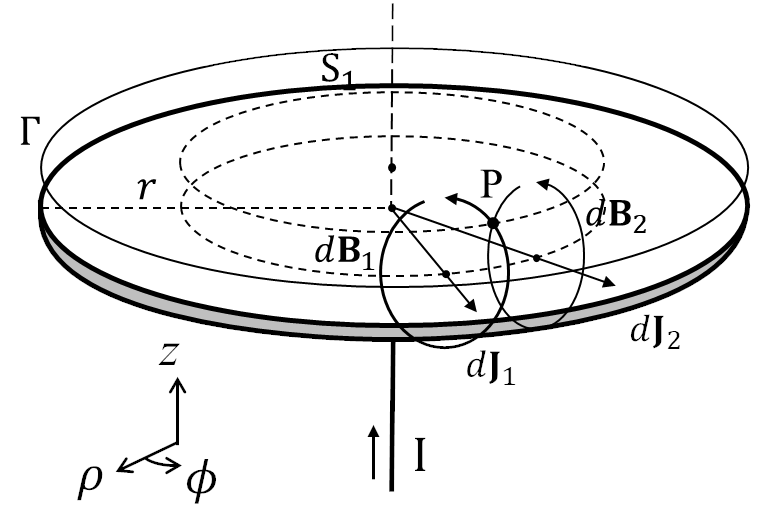}
\caption{Contribution of the radial current in the bottom plate of the capacitor in Fig. \ref{fig:Fig1} to the magnetic field at a point P on surface $\mathrm{S}_1$ when the current is directed outwards. Current elements such as $d\bJ_1$ and $d\bJ_2$ can be always found in pairs which contribute the magnetic field elements $d\bB_1$ and $d\bB_2$ of the same magnitude to $\bB$ at P. Because of the directions of these, the $z$ components cancel and the azimuthal component remains.}
\label{fig:FigA1}
\end{figure}

The Coulombic electric field $\bEC$ and the associated displacement current density on $\mathrm{S}_1$ in FIG. 1 is perpendicular to the capacitor plates and $\mathrm{S}_1$, or 
\beq
 \bEC = E_\mathrm{C} \be_z,           \label{eq:AEC}
\eeq
where $\be_z$ is the unit vector for the $z$ component of the cylindrical coordinate system $(\rho, \phi, z)$. This is because of the symmetric distribution of the electric charge of opposite signs on the surface of the facing plates. It is shown below that the same is true for the induced electric ﬁeld $\bEI$ and the associated displacement current density, or 
\beq
 \bEI = E_\mathrm{I} \be_z         \label{eq:AEI}
\eeq
 on $\mathrm{S}_1$.
 The electric field $\bE$ whose composition is given by (1) satisfies the Amp\`{e}re-Maxwell law
\beq
 \nabla \times \bB = \permitt \permea \frac{\partial \bE}{\partial t }  \label{eq:AAM}
\eeq 
in the area where conduction current is absent. Suppose that $\bB$ has only a $\phi$ component,
\beq
 \bB = B \be_\phi,         \label{eq:AB}
\eeq
 on $\mathrm{S}_1$ just  as the result of the ordinary calculation in the quasistatic approximation. Then, since 
\beq
 \nabla \times \be_\phi = 2 \be_z,   \label{eq:Arotephi}         
\eeq
$\bE$ has only a $z$ component, or
\beq
 \bE = E \be_z,           \label{eq:AE}
\eeq
on $\mathrm{S}_1$. Therefore, considering Eqs. (\ref{eq:E}), (\ref{eq:AEC}), and (\ref{eq:AE}) yields Eq. (\ref{eq:AEI}). In fact, Eq. (\ref{eq:AB}) can be simply shown by focusing on the symmetry of the conduction current that is the source of $\bB$. The Biot-Savart law shows that a linear current on the $z$ axis generates an azimuthal magnetic field expressed as Eq. (\ref{eq:AB}) regardless of the length of the current. Thus the magnetic field generated by  the current in the leads is azimuthal. Also azimuthal is the magnetic field generated by the radial current in the capacitor plates, as seen from Fig. \ref{fig:FigA1}. 
Consider the magnetic field element $d\bB_1$ generated at a point P on $\mathrm{S}_1$ by a radial current element $d\bJ_1$ in the bottom plate. We can always find a current element $d\bJ_2$ whose contribution $d\bB_2$ to the magnetic field at P is of the same magnitude as $d\bB_1$, such that it cancels out the $z$ component of $d\bB_1$ while adds to the $\phi$ component. The same is true for the contribution of the radial current in the top plate to the magnetic field on $\mathrm{S}_1$.  Thus we may conclude that the magnetic field between the plates is expressed as Eq. (\ref{eq:AB}) and therefore $\bE_{\mathrm{I}}$ has only a $z$ component on $\mathrm{S}_1$ as Eq. (\ref{eq:AEI}).

Note: The radial current density  $d\bJ=d\bJ(\rho)$ in the capacitor plate is a decreasing function of the radius $\rho$.\cite{Bartlett90, Milsom20} The magnetic field element $d\bB$ shown in Fig. \ref{fig:FigA1} is the contribution from the whole $d\bJ$ for $0<\rho<r$. The azimuthal direction of the contribution of $d\bJ$ to the magnetic field between the plates is in the opposite direction to that of the contribution of the current in the leads, which makes the magnetic field between the plates for $\rho<r$ quite different from and smaller than the field outside (above or below) the capacitor.

An alternative proof of Eq. (\ref{eq:AEI}) is also possible. Differentiating the left-hand side of Eq. (\ref{eq:AAM}) by $t$ and considering Faraday's law and Eqs. (\ref{eq:irrot}) and (\ref{eq:divfree}) yields
\beq
 \nabla \times \frac{\partial \bB}{\partial t} = - \nabla \times (\nabla \times \bE) = -\nabla \times (\nabla \times \bEI) = -\nabla (\nabla \cdot \bEI )+\nabla^2 \bEI = \nabla^2 \bEI. 
\eeq
From Eqs. (\ref{eq:AB}) and (\ref{eq:Arotephi}), the left-hand side of this equation only has a $z$ component, and so does the right-hand side. Consequently, the $x$ and $y$ components of $\bEI$ satisfy Laplace's equations:
\beq
 \nabla^2 \bE_{\mathrm{I}x} = 0,  \quad     \nabla^2 \bE_{\mathrm{I}y} = 0,      
\eeq
and so they have no local minima or maxima. Furthermore, since they vanish far out on the extension of $\mathrm{S}_1$, they must be zero everywhere on  $\mathrm{S}_1$.

\begin{acknowledgments}
The author gratefully acknowledges Dr. Yutaka Shimomura and Dr. Nazrene Zafar for valuable discussions. Valuable suggestions from the reviewers are also highly appreciated.

\end{acknowledgments}


\begin{thebibliography}{99}
\bibitem{Thompson89} S. P. Thompson, ``On the Magnetic Action of Displacement-currents in a Dielectric,'' Proc. Roy. Soc. A \textbf{45}, 392-393 (1889).
\bibitem{Rizzotto99} R. G. Rizzotto, ``Visualizing Displacement Current - A Classroom Experiment,'' The Physics Teacher,  \textbf{37}, 398 (1999).
\bibitem{BartlettCorle85} D. F. Bartlett and T. R. Corle, ``Measuring Maxwell's Displacement Current Inside a Capacitor'', Phys. Rev. Lett. \textbf{55}, 59-62 (1985).
\bibitem{Hyodo22} T. Hyodo, ``Maxwell's displacement current and the magnetic field between capacitor electrodes,'' Eur. J. Phys. \textbf{43}, 065202 (2022).
\bibitem{Bartlett90} D. F. Bartlett, ``Conduction current and the magnetic field in a circular capacitor,'' Am. J. Phys. \textbf{58}, 1168-1172 (1990). 
\bibitem{Milsom20} J. A. Milsom, ``Untold secrets of the slowly charging capacitor,'' Am. J. Phys. \textbf{88}, 194-199 (2020). 

\end{thebibliography}
\end{document}